\documentclass[preprint]{aastex631}

\shorttitle{Photometry of V1674 Her from PGIR's Fast Readout Mode}
\shortauthors{Hansen}
\graphicspath{{./}{figures/}}

\begin{document}

\title{Second Timescale Photometry of the Very Fast Nova V1674 Her with Palomar Gattini-IR}

\author{Kylie Y. Hansen}
\affiliation{Kavli Institute for Astrophysics and Space Research, Massachusetts Institute of Technology, 70 Vassar St, Cambridge, MA 02139, USA}

\author{Kishalay De}
\affiliation{Kavli Institute for Astrophysics and Space Research, Massachusetts Institute of Technology, 70 Vassar St, Cambridge, MA 02139, USA}
\affiliation{Cahill Center for Astrophysics, California Institute of Technology, 1200 E. California Blvd. Pasadena, CA 91125, USA}

\author{Michael C. B. Ashley}
\affiliation{School of Physics, University of New South Wales, Sydney NSW 2052, Australia}

\author{Mansi M. Kasliwal}
\affiliation{Cahill Center for Astrophysics, California Institute of Technology, 1200 E. California Blvd. Pasadena, CA 91125, USA}

\author{Alexander Delacroix}
\affiliation{Caltech Optical Observatories, California Institute of Technology, Pasadena, CA 91125, USA}

\author{Tim Greffe}
\affiliation{Caltech Optical Observatories, California Institute of Technology, Pasadena, CA 91125, USA}

\author{David Hale}
\affiliation{Caltech Optical Observatories, California Institute of Technology, Pasadena, CA 91125, USA}

\author{Matthew J. Hankins}
\affiliation{Cahill Center for Astrophysics, California Institute of Technology, 1200 E. California Blvd. Pasadena, CA 91125, USA}

\author{Ryan Lau}
\affiliation{Institute of Space \& Astronautical Science, Japan Aerospace Exploration Agency, 3-1-1 Yoshinodai, Chuo-ku, Sagamihara, Kanagawa 252-5210, Japan}

\author{Chengkui Li}
\affiliation{Key Laboratory of Particle Astrophysics, Institute of High Energy Physics, Chinese Academy of Sciences, 19B Yuquan Road, Beijing
100049, China}

\author{Daniel McKenna}
\affiliation{Caltech Optical Observatories, California Institute of Technology, Pasadena, CA 91125, USA}

\author{Anna M. Moore}
\affiliation{Research School of Astronomy and Astrophysics, Australian National University, Canberra, ACT 2611, Australia}

\author{Eran O. Ofek}
\affiliation{Department of Particle Physics \& Astrophysics, Weizmann Institute of Science, Rehovot 76100, Israel}

\author{Roger M. Smith}
\affiliation{Caltech Optical Observatories, California Institute of Technology, Pasadena, CA 91125, USA}

\author{Jamie Soon}
\affiliation{Research School of Astronomy and Astrophysics, Australian National University, Canberra, ACT 2611, Australia}

\author{Roberto Soria}
\affiliation{National Astronomical Observatories, Chinese Academy of Sciences, Beijing 100012, China}
\affiliation{Sydney Institute for Astronomy, The University of Sydney, Sydney, NSW 2006, Australia}

\author{Gokul P. Srinivasaragavan}
\affiliation{Department of Astronomy, University of Maryland, College Park, MD 20742, USA}

\author{Tony Travouillon}
\affiliation{Research School of Astronomy and Astrophysics, Australian National University, Canberra, ACT 2611, Australia}



\begin{abstract}

We report second-timescale infrared photometry of the nova V1674 Her using Palomar Gattini-IR. These observations constitute the first infrared and highest temporal resolution data (resolution of $\approx$ 0.84 s) of the nova reported to date. PGIR observed in this fast readout mode for more than an hour on three nights between 3 and 6 days after discovery. We searched for periodic variability using a Lomb–Scargle periodogram and did not detect anything down to a three sigma upper limit of 0.093 mag. This suggests that the periodic variability detected in the nova by \citealt{J.Patterson} was lower by at least a factor of about 1.65 in the first week of the eruption. 

\end{abstract}



\section{Introduction} \label{sec:intro}
V1674 Her (Nova Her 2021) was discovered by Seiji Ueda (Kushiro, Hokkaido, Japan) on 2021 June 12.5484 UT at an unfiltered CCD magnitude of 8.4 mag. According to observations reported to AAVSO, V1674 Her rose to reach a peak visual brightness of $V \approx 6$ mag on 2021 June 12.9 UT before fading rapidly at a rate of $\sim$2 mags in approximately a day. This nova has sparked interest due to its short $t_2$ timescale (the time to decay by 2 visual magnitudes from peak brightness) of about 1.2d, leading to studies in multiple wavelengths  (e.g., \citealt{quimby}, \citealt{K.L.Page}, \citealt{C.E.Woodward}, \citealt{K.Sokolovsky}, \citealt{E.Aydi}, \citealt{U.Munari}).

In this \emph{Research Note} we report the first infrared photometric observations of V1674 Her. Our data are the highest temporal resolution reported to date, obtained with Palomar Gattini-IR (PGIR). Located at Palomar Observatory, PGIR uses a small, 30-cm, J-band telescope with a field of view of 25 square degrees and a pixel scale of 8.7\arcsec. Under normal operations, PGIR is capable of imaging about 7500 square degrees every night to a median depth of about $J = 15.7$ AB mag outside the Galactic plane (\citealt{gattini}, \citealt{gattini2}). A specialized fast readout mode allows for rapid (exposure time $\approx$ 0.84 s) and continuous (duty cycle $\approx$ 100\%) exposures, allowing one to study very fast time domain phenomena which change over timescales as short as a few seconds (\citealt{FRM}). 


\section{Observations and Data Reduction} \label{sec:observations}

Observations of V1674 Her were taken using the fast readout mode of Palomar Gattini-IR. The source was placed in the lower left quadrant of the detector which provides the best image quality and sensitivity (\citealt{gattini}). Observations were acquired on 2021 June 15, 16, and 18 UT. 7921, 5454, and 5130 images were taken on the three nights respectively.

The data were reduced by performing dark subtraction, flat-fielding, astrometry, and photometry as described in \citealt{FRM}. We performed aperture photometry using a radius of 3 pixels, as well as an annulus radius and width of 11.5 pixels and 3.5 pixels respectively in order to avoid contamination from background sources. We also attempted PSF photometry at the position of the nova by multiplying the best fit PSF model with the reduced image. However, the derived photometry appeared to show residuals correlated with the fractional pixel location of the nova. These residuals were found to arise out of time-varying partial leakage of the nova flux into neighboring pixels as the source drifted across the detector due to imperfect tracking of the telescope. These residuals were not observed in aperture photometry since aperture photometry sums up all of the flux in a conservatively large radius.

\section{Results} \label{sec:results}

\begin{figure}[ht!]
\plotone{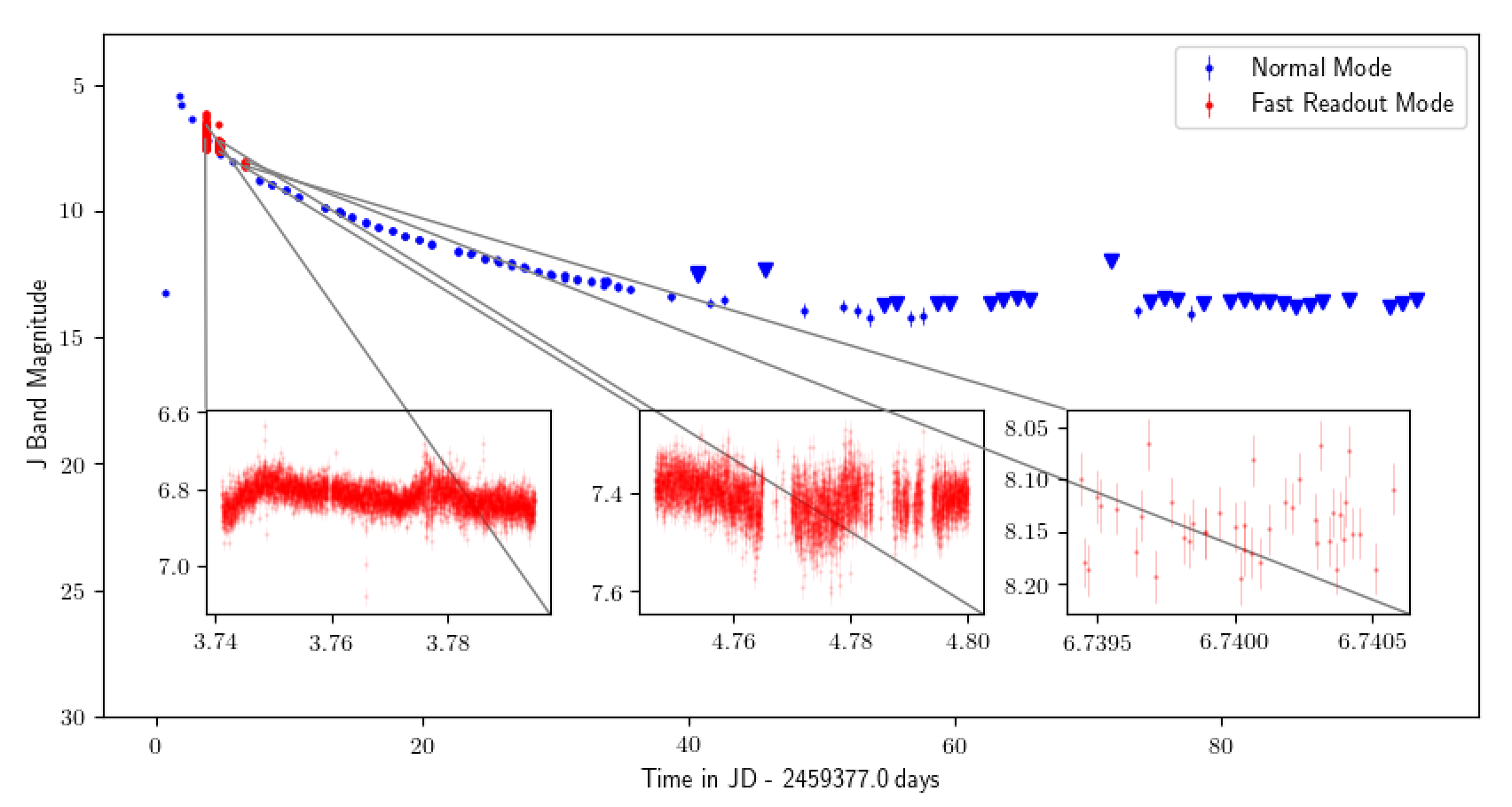}
\caption{The light curve of V1674 Her based solely on data taken from PGIR observations. The blue points mark the data taken during PGIR's normal operations while the red points mark the data taken using PGIR's Fast Readout Mode, which are highlighted in the inserts. The left inset highlights data taken on UTC 2021-06-15, the middle inset highlights data taken on UTC 2021-06-16, and the right inset highlights data taken on UTC 2021-06-18. The arrows represent upper limits of data with signal-to-noise ratios lower than 3.0. \label{fig:lightcurve}}
\end{figure}

Figure \ref{fig:lightcurve} shows our PGIR light curve for V1674 Her, including data from both the normal survey operations and fast readout mode. The insets showcase the second timescale resolution achieved by the fast readout mode. We have removed periods of excess noise caused by the passage of artificial satellites and blurring caused by small telescope slews during the observation. A rise of about 0.1 magnitude can be seen in the left inset, the cause of which is uncertain. This rise was not detected in any random test stars surrounding the nova.

We computed a Lomb-Scargle periodogram of the data on each of the three epochs to search for possible periodic variability. We did not find evidence of a significant frequency peak in the periodogram. \citealt{J.Patterson} detected variability with a period of 8.4 minutes, consistent with the period reported in the archival light curve of the quiescent counterpart by \citealt{ZTF}. We do not detect variability with this period in our analysis. However, \citealt{J.Patterson} reported that the amplitude of the oscillations increased with time after the outburst, and our observations constitute the earliest fast photometry of this source (within a week of the eruption). We attempted to fold our data at the 8.4 minute period and derived a three sigma upper limit on the variability of 0.093 mag. 

\section{Conclusion} \label{sec:conclusion}
We report photometry on the order of seconds of V1674 Her using PGIR. We obtained a total 18505 images of the nova over the course of three nights. We searched for periodic variability of the source using a Lomb-Scargle periodogram, and we did not detect anything down to a three sigma upper limit of 0.093 mag. This suggests that the periodic variability detected in the nova by \citealt{J.Patterson} was lower by at least a factor of about 1.65 in the first week of the eruption. Thus, we conclude that high time resolution observations of novae over longer than a month's baseline are needed to uncover any possible underlying periodicity. 


\bibliography{sample631}{}

\begin{thebibliography}{}
\expandafter\ifx\csname natexlab\endcsname\relax\def\natexlab#1{#1}\fi
\providecommand{\url}[1]{\href{#1}{#1}}
\providecommand{\dodoi}[1]{doi:~\href{http://doi.org/#1}{\nolinkurl{#1}}}
\providecommand{\doeprint}[1]{\href{http://ascl.net/#1}{\nolinkurl{http://ascl.net/#1}}}
\providecommand{\doarXiv}[1]{\href{https://arxiv.org/abs/#1}{\nolinkurl{https://arxiv.org/abs/#1}}}

\bibitem[{{Aydi} {et~al.}(2021){Aydi}, {Sokolovsky}, {Chomiuk}, {Strader},
  {Kawash}, {Page}, {Boussin}, {Ikonnikova}, {Atapin}, {Belinski}, {Burlak},
  {Dodin}, {Maslennikova}, {Postnov}, {Potanin}, {Safonov}, {Shatsky},
  {Tatarnikov}, {Korotkiy}, {Stanek}, {Kochanek}, \& {Shappee}}]{E.Aydi}
{Aydi}, E., {Sokolovsky}, K.~V., {Chomiuk}, L., {et~al.} 2021, The Astronomer's
  Telegram, 14710, 1

\bibitem[{{De} {et~al.}(2020{\natexlab{a}}){De}, {Hankins}, {Kasliwal},
  {Moore}, {Ofek}, {Adams}, {Ashley}, {Babul}, {Bagdasaryan}, {Burdge},
  {Burnham}, {Dekany}, {Declacroix}, {Galla}, {Greffe}, {Hale}, {Jencson},
  {Lau}, {Mahabal}, {McKenna}, {Sharma}, {Shopbell}, {Smith}, {Soon},
  {Sokoloski}, {Soria}, \& {Travouillon}}]{gattini}
{De}, K., {Hankins}, M.~J., {Kasliwal}, M.~M., {et~al.} 2020{\natexlab{a}},
  \pasp, 132, 025001, \dodoi{10.1088/1538-3873/ab6069}

\bibitem[{{De} {et~al.}(2020{\natexlab{b}}){De}, {Ashley}, {Andreoni},
  {Kasliwal}, {Soria}, {Srinivasaragavan}, {Cai}, {Delacroix}, {Greffe},
  {Hale}, {Hankins}, {Li}, {McKenna}, {Moore}, {Ofek}, {Smith}, {Soon},
  {Travouillon}, \& {Zhang}}]{FRM}
{De}, K., {Ashley}, M. C.~B., {Andreoni}, I., {et~al.} 2020{\natexlab{b}},
  \apjl, 901, L7, \dodoi{10.3847/2041-8213/abb3c5}

\bibitem[{{Moore} \& {Kasliwal}(2019)}]{gattini2}
{Moore}, A.~M., \& {Kasliwal}, M.~M. 2019, Nature Astronomy, 3, 109,
  \dodoi{10.1038/s41550-018-0675-x}

\bibitem[{{Mroz} {et~al.}(2021){Mroz}, {Burdge}, {Roestel}, {Prince}, {Kong},
  \& {Li}}]{ZTF}
{Mroz}, P., {Burdge}, K., {Roestel}, J.~v., {et~al.} 2021, The Astronomer's
  Telegram, 14720, 1

\bibitem[{{Munari} {et~al.}(2021){Munari}, {Valisa}, \&
  {Dallaporta}}]{U.Munari}
{Munari}, U., {Valisa}, P., \& {Dallaporta}, S. 2021, The Astronomer's
  Telegram, 14704, 1

\bibitem[{{Page} {et~al.}(2021){Page}, {Orio}, {Sokolovsky}, \&
  {Kuin}}]{K.L.Page}
{Page}, K.~L., {Orio}, M., {Sokolovsky}, K.~V., \& {Kuin}, N.~P.~M. 2021, The
  Astronomer's Telegram, 14747, 1

\bibitem[{{Patterson} {et~al.}(2021){Patterson}, {Epstein-Martin},
  {Vanmunster}, \& {Kemp}}]{J.Patterson}
{Patterson}, J., {Epstein-Martin}, M., {Vanmunster}, T., \& {Kemp}, J. 2021,
  The Astronomer's Telegram, 14856, 1

\bibitem[{Quimby {et~al.}(2021)Quimby, Shafter, \& Corbett}]{quimby}
Quimby, R.~M., Shafter, A.~W., \& Corbett, H. 2021, The Detailed Light Curve
  Evolution of V1674 Her (Nova Her 2021).
\newblock \doarXiv{2107.05763}

\bibitem[{{Sokolovsky} {et~al.}(2021){Sokolovsky}, {Aydi}, {Chomiuk}, {Kawash},
  {Strader}, {Babul}, {Sokoloski}, {Linford}, {Mukai}, \& {Li}}]{K.Sokolovsky}
{Sokolovsky}, K., {Aydi}, E., {Chomiuk}, L., {et~al.} 2021, The Astronomer's
  Telegram, 14731, 1

\bibitem[{{Woodward} {et~al.}(2021){Woodward}, {Banerjee}, {Wagner},
  {Starrfield}, \& {Evans}}]{C.E.Woodward}
{Woodward}, C.~E., {Banerjee}, D.~P.~K., {Wagner}, R.~M., {Starrfield}, S., \&
  {Evans}, A. 2021, The Astronomer's Telegram, 14728, 1

\end{thebibliography}
\bibliographystyle{aasjournal}



\end{document}